\documentclass{article}
\usepackage{spconf,amsmath,graphicx,hyperref,booktabs, bbding, pifont}
\usepackage{balance, multirow}

\title{DiaScriber: A Speech LLM for Joint Diarization and Transcription \\in Multi-Speaker Scenarios}
\name{
\parbox{\linewidth}{\centering
Bingshen Mu$^{1*}$, Xian Shi$^{2*}$, Xiong Wang$^2$, Zhifang Guo$^2$, Ting He$^2$, Xize Cheng$^2$, Yu Xi$^2$, Jin Xu$^2$, Lei Xie$^1$}\thanks{$^*$Equal Contribution.}}
\address{$^1$Audio, Speech and Language Processing Group (ASLP@NPU), School of Computer Science,\\ Northwestern Polytechnical University, Xi’an, China\\$^2$Independent Researcher}

\begin{document}
\ninept
\maketitle
\begin{abstract}
Multi-speaker automatic speech recognition (MSASR) aims to jointly predict content transcriptions, speaker identities, and timestamps, thereby addressing the key question of "who spoke what and when" and holds substantial practical value in real-world multi-speaker scenarios. However, MSASR still encounters considerable challenges in the presence of fast turn transitions, overlapping speech, and complex, diverse multi-speaker scenarios.
In this work, we propose DiaScriber, an end-to-end multi-speaker diarization and transcription model built on a speech large language model. We first construct diverse data pipelines to cover a wide variety of multi-speaker scenarios and their complexities, including validation and refinement, turn-transition and overlapping-speech simulation, and multimodal annotation.
Furthermore, DiaScriber is developed based on the pretrained version of Qwen3.5-Omni through a three-stage training strategy involving continual pretraining, supervised fine-tuning, and reinforcement learning.
Experiments show that DiaScriber achieves superior performance over comparison methods across extensive multi-speaker scenario test sets and demonstrates outstanding generalization ability in unseen multi-speaker scenarios.
\end{abstract}
\vspace{-4.5pt}
\section{Introduction}
\vspace{-4.5pt}
Automatic speech recognition (ASR) generally focuses on predicting content transcriptions for single-speaker speech~\cite{qwen2026qwenasr, fireredasr, Radford2023Robust}; in contrast, multi-speaker ASR (MSASR) with speaker diarization predicts content transcriptions annotated with speaker identities and timestamps, thereby addressing the critical question of ``who spoke what and when"~\cite{vibevoiceasr, donghua2026moss, dai2026soulx, dai2026joint, lin2026speaker}. By providing a more comprehensive transcription of real-world human communication, MSASR is of substantial value in interactive settings including meetings, interviews, and classrooms. 
However, real-world MSASR still encounters notable challenges. Specifically, multi-speaker interactions commonly involve fast turn transitions, long-term dependencies, varying acoustic conditions, and frequent overlapping speech, all of which significantly increase the difficulty of MSASR. Additionally, many practical applications further require MSASR to maintain stable performance across different scenarios, including meetings, classrooms, interviews, podcasts, and open-domain dialogues, where speaker counts, interaction styles, recording setups, and noise conditions can differ significantly.

Traditional MSASR typically employs a cascaded processing pipeline, where speaker diarization, speech segmentation, and ASR are assigned to independent modules~\cite{fan2022m2met, mu2026summary, he2026survey}. Although this design provides strong modular replaceability and engineering flexibility, its limitations are equally evident. Individual modules are optimized for separate objectives, preventing them from fully capturing the inherent dependencies among speaker identities, timestamps, and content. Moreover, errors in upstream modules propagate through subsequent stages, causing significant error accumulation, particularly under conditions with overlapping speech or frequent speaker changes. Finally, cascaded pipelines depend heavily on complex post-processing and heuristic rules to merge outputs from multiple modules, limiting both end-to-end optimization and deployment across diverse scenarios.

Recent achievements of speech large language models (LLMs) in tasks such as ASR and timestamp prediction have provided a promising direction for the further development of MSASR~\cite{qwen2026qwenasr, mu2026llm}.
Early work focuses on semi-cascaded pipelines, which maintain independent speaker diarization and speaker verification modules, subsequently inputting the extracted speaker identities and timestamps into speech LLMs for global processing to improve MSASR performance~\cite{li2026dm}. Although semi-cascaded pipelines enhance global consistency to some extent, they still depend on the outputs of preceding modules and thus fail to overcome the error propagation and module incompatibility problems of traditional cascaded pipelines.
Existing end-to-end MSASR methods based on speech large language models have achieved encouraging performance, yet some limitations persist. SoulX-Transcriber~\cite{dai2026soulx} supports no more than 10 minutes of input and cannot process longer speech. VibeVoice-ASR~\cite{vibevoiceasr} and MOSS-Transcribe-Diarize~\cite{donghua2026moss} are capable of processing speech inputs longer than one hour, but their stability across complex and diverse scenarios still requires further improvement.

In this work, we propose \textbf{\textit{DiaScriber}}, an end-to-end MSASR model based on speech LLMs that simultaneously predicts diarization and transcription across diverse, complex multi-speaker scenarios.
First, we carefully construct diverse data pipelines to provide DiaScriber with data sources that cover a wide range of multi-speaker scenarios and complexities of human communication.
Specifically, the data validation and refinement pipeline aims to verify the quality of timestamp and content annotations in internal multi-speaker speech data and revise the low-quality portions, covering scenarios such as dialogues, podcasts, interviews, movies, and commentary in Mandarin, multiple Chinese dialects, and English.
The data simulation pipeline consists of two parts. One uses reference-guided voice cloning to extend the speech of multiple target speakers, which are then concatenated into long speech with randomized intervals and orders to enrich turn transitions. The other improves the diversity of overlapping speech by cropping and shifting adjacent segments from different speakers in existing long speech.
The multimodal data annotation pipeline combines multiple ASR outputs, speaker diarization, visual face tracks, audio-visual consistency, and the annotation capabilities of commercial multimodal LLM APIs to generate structured annotations with speaker identities, timestamps, and content transcriptions from in-the-wild multi-speaker videos.
Moreover, DiaScriber is developed from the pretrained version of Qwen3.5-Omni~\cite{team2026qwen3} through a three-stage training strategy consisting of continual pretraining (CPT), supervised fine-tuning (SFT), and reinforcement learning (RL).
Extensive experiments demonstrate the effectiveness of our diverse data pipeline and three-stage training strategy. Across multiple multi-speaker test sets, DiaScriber achieves significantly better diarization error rate (DER), concatenated minimum-permutation word error rate (cpWER), and time-constrained minimum-permutation word error rate (tcpWER) than VibeVoice-ASR and MOSS-Transcribe-Diarize. In addition, DiaScriber shows outstanding generalization performance, delivering the best DER, cpWER, and tcpWER on unseen multi-speaker scenarios.
\vspace{-4.5pt}
\section{Data Pipelines}
\vspace{-4.5pt}
\subsection{Validation and Refinement Pipeline}
\vspace{-4.5pt}
The validation phase aims to determine the quality of existing manually annotated long-form multi-speaker speech data in terms of content transcriptions, timestamps, and speaker identities.
Specifically, for each long-form multi-speaker speech, we extract several speech segments from the beginning, middle, and end according to the manually annotated timestamps. These segments are then transcribed by Qwen3-ASR-Flash~\cite{qwen2026qwenasr} to produce hypotheses, after which we calculate the word error rate (WER) and the boundary consecutive error rate (BCER) between the hypotheses and the manual transcriptions. Notably, BCER measures the consecutive insertion and deletion error rates at the beginning and end of each segment. 
We believe that a higher BCER indicates less accurate manual timestamp annotation because timestamp shifts are reflected in the ASR hypotheses at the boundaries of speech segments as increased insertion and deletion errors relative to the manually annotated transcriptions.
Subsequently, we compute the speaker similarity among all speech segments belonging to a specific speaker in the long-form speech to validate the accuracy of the manually annotated speaker identities.

Data exceeding the predefined thresholds are used directly for subsequent training, while the remaining data proceed to the refinement phase.
For multi-speaker long-form speech with high BCER, since refining the timestamps is both expensive and difficult, we keep the low-quality manually annotated timestamps unchanged and instead refine the manually annotated transcriptions for those speech segments by combining the ASR hypotheses at the segment boundaries with the manually annotated transcriptions for the non-boundary portions.
For multi-speaker long-form speech with low-quality manually annotated speaker identities, we divide the entire speech into segments based on timestamps, obtain a speaker embedding for each segment, and perform reclustering to produce repaired speaker-identity annotations.
The validation and refinement pipeline preserves as much manually annotated multi-speaker long-form speech as possible, covering dialogues, podcasts, interviews, films, and commentary scenarios in English, Mandarin, and dialects.
\vspace{-18pt}
\subsection{Simulation Pipeline}
\vspace{-4.5pt}
This pipeline consists of two components: synthesis–concatenation and crop–shift.
In synthesis–concatenation, we first employ Qwen3-TTS~\cite{hu2026qwen3} for reference-based voice cloning. For each target speaker, we provide a reference speech sample as the speaker prompt and a text sequence to be synthesized; Qwen3-TTS then generates a speech segment that preserves the target speaker's voice characteristics and contains the specified text.
Notably, texts used for synthesis are not required to maintain semantic continuity across adjacent segments.
Once we obtain real and synthesized speech segments from multiple target speakers, we concatenate them in random order with random time intervals to construct multi-speaker long-form speech, increasing the diversity of turn transitions.
To make the concatenated long-form speech more similar to real-world acoustic conditions, we apply background noise and reverberation perturbations to simulate characteristics under different acoustic environments.

In crop–shift, given existing multi-speaker long-form speech, we select adjacent boundary points between different speakers based on the annotations. We use each such boundary point to split the original long-form speech into two segments. We shift the latter segment forward along the timeline by a random duration and concatenate it with the former segment again. Since the start time of the latter speaker is advanced, the two segments overlap near the boundary, resulting in more challenging two-speaker overlapping-speech long-form samples. By controlling the randomness of the shift magnitude, we can further cover two-speaker overlap patterns with different durations and ratios.
We further support extracting additional segments from speech attributed to other speakers, shifting them along the timeline, and inserting them into existing overlapping regions, thereby extending the original two-speaker overlap to samples involving three or even more overlapping speakers.
The multi-speaker long-form speech samples produced by synthesis–concatenation and crop–shift components of the simulation pipeline do not emphasize semantic continuity across adjacent segments. Instead, they are designed to provide targeted augmentation for two key challenges in the MSASR task: diverse turn transitions and overlapping speech.
\begin{figure}[t]
\centering
\includegraphics[width=0.8\columnwidth]{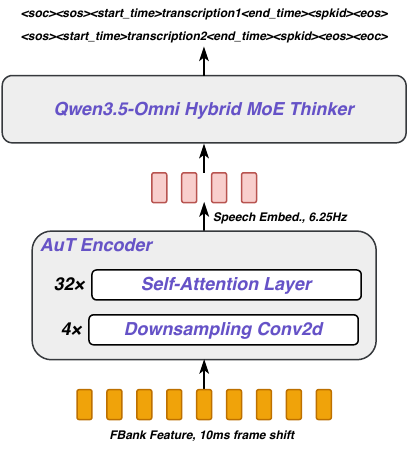}
\caption{Overview of DiaScriber.}
\vspace{-9pt}
\label{fig1}
\end{figure}
\vspace{-4.5pt}
\subsection{Multimodal Annotation Pipeline}
\vspace{-4.5pt}
This pipeline is intended to produce annotations with speaker identities, timestamps, and content transcriptions from real-world multi-speaker in-the-wild media.
We first process the audio stream in the media. Through voice activity detection~\cite{SileroVAD}, speaker diarization~\cite{Bredin2023pyannote}, and ROVER~\cite{fiscus1997post} fusion of multiple ASR predictions, we obtain preliminary content transcriptions along with the corresponding speaker identities and timestamps.
After that, we split long-form media based on the preliminary speaker diarization results. 

Next, for each media segment's video stream, we perform frame-level face detection and cross-frame association to extract continuous face tracks. Each face track corresponds to a candidate visual character that appears continuously over time. We then treat face tracks as the basic units for character-level clustering, assigning stable pseudo-identity labels to the same person within a given media segment. 
To further establish the correspondence between visual characters and speech segments, we employ SyncNet~\cite{Chung2016out} to calculate an audio–visual consistency score between each face track and the speech within its corresponding time range. 
By incorporating audio–visual consistency analysis, the multimodal annotation pipeline can leverage visual information to improve the reliability of speaker discrimination and alignment in situations where relying solely on speaker diarization may lead to speaker fragmentation or identity confusion.
By combining face-track clustering results with speaker diarization results, we merge speech segments that belong to the same visual character but were assigned different speaker identities. 
We then feed the media segment and preliminary multimodal annotations into Gemini-3.1-pro-preview, guiding it to perform fine-grained understanding of the current media segment in multimodal context and output speaker identities, timestamps, and corresponding content transcriptions.
The multimodal annotation pipeline primarily annotates media data from interviews, movies, and television drama scenarios.
\vspace{-4.5pt}
\section{DiaScriber}
\vspace{-4.5pt}
\subsection{Overall Architecture}
\vspace{-4.5pt}
DiaScriber is built on the pretrained version of Qwen3.5-Omni and comprises a speech encoder, a linear projector layer, and an LLM backend.
The speech encoder in DiaScriber is based on the Audio Transformer (AuT), trained on 40 million hours of speech–text pairs from Qwen3-ASR. It exhibits strong general speech representation capabilities, providing robust acoustic and semantic priors for multi-speaker scenarios.
The Fbank speech features are downsampled 16 times using 4 Conv2D blocks, then fed into self-attention layers to produce speech embeddings at a 6.25Hz frame rate. The low-frame-rate speech representations of AuT improve efficiency for long-form inputs, laying the foundation for multi-speaker long-form speech processing. However, the excessively low frame rate in speech representations may also adversely affect the accuracy of timestamp predictions.
To balance potential streaming inference requirements with short chunks and offline inference with entire long-form speech, DiaScriber adopts a flash attention window size ranging from 1s to 8s during training within AuT.
Subsequently, AuT provides the speech embedding sequence to the LLM backbone, which generates a target sequence that includes speaker identity, timestamps, and transcription content in the following format: \textit{\textless soc\textgreater\allowbreak \textless sos\textgreater\allowbreak \textless start\_time\textgreater\allowbreak transcription1\textless end\_time\textgreater\allowbreak \textless spkid\textgreater\allowbreak \textless eos\textgreater\allowbreak\textless sos\textgreater\allowbreak\textless start\_time\textgreater\allowbreak transcription2\textless end\_time\textgreater\allowbreak\textless spkid\textgreater\allowbreak\textless eos\textgreater\allowbreak\textless eoc\textgreater\allowbreak}.
In this format, \textit{\textless soc\textgreater\allowbreak} and \textit{\textless eoc\textgreater\allowbreak} represent the special tokens for the start and end of the conversation, while \textit{\textless sos\textgreater\allowbreak} and \textit{\textless eos\textgreater\allowbreak} denote the special tokens for the start and end of the speech segment. Additionally, \textit{\textless start\_time\textgreater\allowbreak} and \textit{\textless end\_time\textgreater\allowbreak} indicate the start and end timestamps of the speech segment, and \textit{\textless spkid\textgreater\allowbreak} represents the speaker identity.
\begin{table}[t]
    \caption{Details of the internal test sets, where ``OOD" indicates whether the test set is out of domain.}
    \label{tab:tabel1}
    \centering
\scalebox{0.9}{
\begin{tabular}{lcccc}
\toprule
\textbf{ID} & \textbf{Lang.} & \textbf{Spk.} & \textbf{Scenarios}                                                                             & \textbf{OOD} \\ \midrule
Internal-1  & ZH             & 2$\sim$5      & \begin{tabular}[c]{@{}c@{}}Dialectal dialogue, podcast,\\ interview, movie\end{tabular}        & \ding{55}        \\
Internal-2  & EN             & 2$\sim$4      & \begin{tabular}[c]{@{}c@{}}Podcast, interview,\\ commentary\end{tabular}                       & \ding{55}         \\
Internal-3  & ZH             & 2$\sim$3      & Non-overlapped dialogue                                                                        & \ding{51}         \\
Internal-4  & ZH             & 2$\sim$3      & Overlapped dialogue                                                                            & \ding{51}         \\
Internal-5  & ZH             & 2             & \begin{tabular}[c]{@{}c@{}}Adult-child dialogue,\\ child-child dialogue\end{tabular}           & \ding{51}         \\
Internal-6  & ZH/EN          & 2$\sim$5      & \begin{tabular}[c]{@{}c@{}}Podcast, interview, news,\\ commentary, live-streaming\end{tabular} & \ding{51}         \\
Internal-7  & ZH             & 2             & Thematic dialogue                                                                              & \ding{55}         \\
Internal-8  & ZH             & 2$\sim$5      & Real-time online voice chat                                                                    & \ding{55}         \\ \bottomrule
\end{tabular}}
\vspace{-9pt}
\end{table}

\begin{table*}[t]
    \caption{DER (\%), cpWER (\%), and tcpWER (\%) results of DiaScriber and comparison methods on the open-source and internal test sets.}
    \label{tab:tabel2}
    \centering
\begin{tabular}{lccccccccc}
\toprule
\multirow{2}{*}{} & \multicolumn{3}{c}{\textbf{VibeVoice-ASR}}      & \multicolumn{3}{c}{\textbf{MOSS-Transcribe-Diarize}} & \multicolumn{3}{c}{\textbf{DiaScriber}}         \\ \cmidrule{2-10} 
                  & \textbf{DER} & \textbf{cpWER} & \textbf{tcpWER} & \textbf{DER}   & \textbf{cpWER}   & \textbf{tcpWER}  & \textbf{DER} & \textbf{cpWER} & \textbf{tcpWER} \\ \midrule
AliMeeting~\cite{fan2022m2met}        & 10.2         & 28.0           & 28.1            & 4.1            & 16.2             & 16.3             & 3.6          & 12.0           & 12.4            \\
AISHELL-4~\cite{fu2021aishell}         & 7.6          & 22.2           & 22.6            & 3.4            & 14.0             & 14.3             & 3.7          & 15.3           & 16.0            \\
MagicData-RAMC~\cite{yang2022open}    & 5.9          & 15.8           & 16.9            & 5.5            & 13.5             & 14.3             & 7.2          & 15.9           & 16.8            \\
AMI~\cite{carletta2005ami}               & 119.6        & 71.7           & 74.9            & 24.0           & 17.7             & 20.4             & 5.5          & 10.3           & 10.8            \\
MLCSLM-EN~\cite{mu2026summary}         & 2.3          & 13.0           & 13.4            & 2.1            & 12.5             & 12.9             & 4.0          & 15.5           & 16.0            \\ \midrule
Internal-1        & 16.6         & 28.2           & 29.3            & 47.3           & 35.2             & 97.2             & 6.1          & 10.8           & 12.8            \\
Internal-2        & 19.7         & 39.1           & 45.4            & 21.6           & 51.5             & 58.8             & 6.9          & 17.9           & 20.7            \\
Internal-3        & 13.2         & 28.7           & 28.7            & 3.8            & 3.9              & 3.9              & 3.1          & 6.6            & 6.8             \\
Internal-4        & 15.8         & 34.0           & 34.3            & 4.2            & 5.1              & 5.2              & 1.3          & 2.2            & 2.3             \\
Internal-5        & 10.6         & 8.2            & 8.2             & 2.6            & 1.9              & 1.9              & 2.8          & 1.6            & 1.6             \\
Internal-6        & 15.8         & 31.0           & 31.9            & 15.6           & 20.3             & 20.9             & 10.4         & 16.8           & 17.5            \\
Internal-7        & 70.7         & 23.6           & 141.9           & 6.3            & 7.2              & 8.6              & 5.4          & 11.9           & 13.9            \\
Internal-8        & 31.7         & 32.0           & 32.7            & 101.6          & 103.4            & 120.5            & 10.8         & 15.8           & 16.1            \\ \midrule
Avg.              & 26.1         & 28.9           & 39.1            & 18.6           & 23.3             & 30.4             & \textbf{5.4} & \textbf{11.8}  & \textbf{12.6}   \\ \bottomrule
\end{tabular}
\vspace{-9pt}
\end{table*}
\vspace{-4.5pt}
\subsection{Training Strategy}
\vspace{-4.5pt}
\label{training-strategy}
The training process of DiaScriber consists of three stages: CPT, SFT and RL.
The objective of the CPT stage is to enable DiaScriber to better adapt to the MSASR task and its output format while leveraging the general speech understanding capabilities of Qwen3.5-Omni.
Specifically, this stage employs large-scale and diverse training data, including open-source data and data generated by the verification and refinement pipeline, the simulation pipeline, and the multimodal annotation pipeline, for a total of 85,000 hours. These data cover a wide range of scenarios, including dialogues, meetings, classrooms, interviews, podcasts, movies, and commentary, and include complex phenomena such as fast turn transitions, overlapping speech, complex noise, and varying numbers of speakers. After the CPT stage, DiaScriber can gradually evolve from its original general speech-understanding capabilities into capabilities better aligned with the MSASR task.
The loss function used during CPT stage is the negative log-likelihood loss under teacher forcing.
The objective of the SFT stage is to improve the stability of DiaScriber's output format and its specialization for the MSASR task, transforming it from a general speech understanding model into an MSASR expert model. Specifically, the SFT stage uses fixed input and output templates to ensure DiaScriber generates diarization and transcription predictions strictly in the prescribed format, while discouraging content irrelevant to MSASR. During the SFT stage, we use open-source data together with carefully selected data produced by the validation and refinement pipeline, covering the same multi-speaker scenarios as those in the CPT stage, for a total of about 4,600 hours.
The loss function used during the SFT stage is consistent with that of the CPT stage.
Although SFT can substantially improve DiaScriber's output quality, its optimization objective remains token-level maximum likelihood rather than MSASR evaluation metrics.
During the RL stage, we employ group sequence policy optimization (GSPO)~\cite{zheng2025gspo} to further enhance DiaScriber's performance on MSASR evaluation metrics.
Specifically, we use three reward components directly aligned with the MSASR objectives, and construct a composite reward based on DER, cpWER, and tcpWER. Through a weighted combination of these three rewards, DiaScriber can simultaneously optimize its capabilities at three levels: ``who is speaking", ``who said what", and ``who said what and when".
Compared with the preceding two stages, RL primarily enhances DiaScriber's performance and stability. For RL training, we leverage 1,000 unseen multi-speaker long-form speeches that cover both dialogue and meeting scenarios.
\vspace{-4.5pt}
\section{Experiments}
\vspace{-4.5pt}
\subsection{Experimental Setup}
\vspace{-4.5pt}
The scale of the training data used by DiaScriber at each stage can be found in section~\ref{training-strategy}.
During the evaluation stage, we use the open-source end-to-end MSASR models VibeVoice-ASR and MOSS-Transcribe-Diarize as comparison methods. The test sets consist of widely used open-source Chinese and English test sets, along with internal test sets covering diverse scenarios. 
DiaScriber and all other methods take raw long-form speech as input during evaluation, without any segmentation.
Table~\ref{tab:tabel1} presents the details of the internal test sets.
DiaScriber’s AuT encoder has 600M parameters, and the LLM backend is a 23B-parameter MoE architecture with 2.6B active parameters.
\vspace{-4.5pt}
\subsection{Experimental Results}
\vspace{-4.5pt}
Table~\ref{tab:tabel2} reports the DER, cpWER, and tcpWER results of DiaScriber and other comparison methods on the open-source and internal test sets. On the open-source test sets, MOSS-Transcribe-Diarize and DiaScriber each have their respective advantages and disadvantages. On the internal test sets, which involve more diverse multi-speaker scenarios, DiaScriber outperforms the others in most cases and achieves substantially lower average metrics regardless of whether the test sets are out of domain.
However, the unusually high metrics obtained by VibeVoice-ASR and MOSS-Transcribe-Diarize on some internal test sets stem from their tendency to generate frequent hallucinations during inference, indicating limited generalization ability.
In our view, even though DiaScriber contains a substantial number of parameters, its low frame rate imposes a limitation on MSASR performance.

Table~\ref{tab:tabel3} shows DiaScriber's average DER, cpWER, and tcpWER results at the CPT, SFT, and RL training stages on the open-source test sets. Across the three-stage training process of DiaScriber, all three metrics consistently decrease, indicating that the three-stage training strategy can steadily improve DiaScriber’s MSASR performance.
The CPT stage only provides DiaScriber with preliminary MSASR capability. In contrast, the SFT stage brings a substantial performance improvement, showing that it can effectively enhance DiaScriber’s ability on the MSASR task. Although the performance gain in the RL stage is smaller than that in the SFT stage, it still demonstrates that RL can further optimize DiaScriber’s prediction quality and help it achieve better overall performance.

Table~\ref{tab:tabel4} indicates that all three reward components contribute positively to DiaScriber's performance, together establishing a highly effective composite reward for the RL stage. Removing the cpWER reward causes the most significant degradation in overall performance, while removing the tcpWER reward primarily affects transcription-related metrics. In contrast, removing the DER reward has a greater impact on speaker discrimination ability and also indirectly weakens transcription performance. These findings show that the three rewards constrain DiaScriber’s behavior from different perspectives and are highly complementary.
Using these rewards together enables a better balance between speaker discrimination and transcription accuracy, leading to better overall performance.
\begin{table}[t]
    \caption{Average DER (\%), cpWER (\%), and tcpWER (\%) results of three different training stages on the open-source test sets.}
    \label{tab:tabel3}
    \centering
\begin{tabular}{lccc}
\toprule
          & \textbf{DER} & \textbf{cpWER} & \textbf{tcpWER} \\ \midrule
CPT Stage & 8.2          & 18.1           & 19.3            \\
SFT Stage & 5.5          & 14.4           & 14.8            \\
RL Stage  & 4.8          & 13.8           & 14.4            \\ \bottomrule
\end{tabular}
\vspace{-9pt}
\end{table}

\begin{table}[t]
    \caption{Average DER (\%), cpWER (\%), and tcpWER (\%) across all test sets for the ablation study of different rewards in the RL stage.}
    \label{tab:tabel4}
    \centering
\begin{tabular}{lccc}
\toprule
                  & \textbf{DER} & \textbf{cpWER} & \textbf{tcpWER} \\ \midrule
RL Stage          & 5.4          & 11.8           & 12.6            \\
w/o DER Reward    & 6.2          & 12.5           & 13.5            \\
w/o cpWER Reward  & 6.3          & 12.4           & 13.7            \\
w/o tcpWER Reward & 5.5          & 13.0           & 13.9            \\ \bottomrule
\end{tabular}
\vspace{-9pt}
\end{table}
\vspace{-4.5pt}
\section{Conclusion}
\vspace{-4.5pt}
In this work, we propose DiaScriber, an end-to-end multi-speaker diarization and transcription model for the MSASR task. We construct multiple data pipelines, including validation and refinement, simulation, and multimodal annotation, to provide DiaScriber with large-scale training corpora covering diverse multi-speaker scenarios and address the key challenges of the MSASR task.
Moreover, DiaScriber is developed based on the pretrained version of Qwen3.5-Omni and converted from a general speech understanding model into an MSASR expert model via a three-stage training strategy involving CPT, SFT, and RL.
Experiments show that DiaScriber outperforms comparison methods across extensive test sets covering diverse multi-speaker scenarios, while also demonstrating strong generalization and stability in unseen multi-speaker environments.

\clearpage
\balance
\bibliographystyle{IEEEbib}
\bibliography{refs}

\end{document}